\documentclass[a4paper]{jpconf}

\usepackage{amssymb}
\usepackage{amsmath}
\usepackage{amscd}
\usepackage{latexsym}

\usepackage{float}

\usepackage{cite}

\usepackage{graphicx}
\usepackage{subfigure}

\newcommand{\be}{\begin{equation}}
\newcommand{\ee}{\end{equation}}

\newcommand{\dlt}{\delta}
\newcommand{\prt}{\partial}
\newcommand{\bfr}{{\bf r}}

\newcommand{\vp}{\varphi}

\newcommand{\al}{\alpha}

\newcommand{\om}{\omega}
\newcommand{\Om}{\Omega}

\begin{document}

\title{Vortex rings and vortex ring solitons in shaken
Bose-Einstein condensate}

\author{V.I. Yukalov$^1$, A.N. Novikov$^{1,2}$, E.P. Yukalova$^3$,
and V.S. Bagnato$^2$}

\address{
$^1$Bogolubov Laboratory of Theoretical Physics,
Joint Institute for Nuclear Research, \\ Dubna 141980, Russia}

\address{
$^2$Instituto de Fisica de S\~{a}o Calros, Universidade de S\~{a}o Paulo,
CP 369, 13560-970 S\~{a}o Carlos, \\ S\~{a}o Paulo, Brazil}

\address{
$^3$Laboratory of Informational Technologies,
Joint Institute for Nuclear Research, \\ Dubna 141980, Russia}

\ead{yukalov@theor.jinr.ru}

\begin{abstract}
In a shaken Bose-Einstein condensate, confined in a vibrating trap, there
can appear different nonlinear coherent modes. Here we concentrate on two 
types of such coherent modes, vortex ring solitons and vortex rings. In a 
cylindrical trap, vortex ring solitons can be characterized as nonlinear
Hermite-Laguerre modes, whose description can be done by means of optimized
perturbation theory. The energy, required for creating vortex ring solitons,
is larger than that needed for forming vortex rings. This is why, at a
moderate excitation energy, vortex rings appear before vortex ring solitons.  
The generation of vortex rings is illustrated by numerical simulations for 
trapped $^{87}$Rb atoms.
\end{abstract}

\section{Introduction}

Nonlinear Schr\"{o}dinger (NLS) equation exhibits different kinds of
soliton-like solutions. Various solitonic solutions are well known for
the NLS equation in nonlinear optics \cite{Kivshar_1,Briedis_2} and in
Bose-Einstein condensates \cite{Komineas_3,Frantzeskakis_4}. The behavior
of solitons in both these nonlinear media possesses many common features
\cite{Malomed_5,Kartashov_6}.

Here we concentrate on two types of solutions to the NLS equation, vortex
ring solitons and vortex rings. The theoretical study of such solutions
is usually based on the assumption of their existence at the initial moment
of time, after which one considers their temporal dynamics starting from
the given initial condition. In a realistic situation, in order to create
a soliton in a Bose-condensed system, it is necessary to apply an external
perturbation rendering the condensate strongly nonequilibrium. There are
several ways of perturbing the system. Trap rotation or laser rotation
inside a trap is known to produce quantum vortices and vortex lattices
\cite{Pethick_7}. To generate a variety of other coherent modes, it has
been suggested \cite{Yukalov_8,Yukalov_9,Yukalov_10} to shake the condensed
system, which can be realized either by trap shaking
\cite{Yukalov_8,Yukalov_9,Yukalov_10,Yukalov_11} or by scattering length
modulation \cite{Yukalov_12,Yukalov_13}. A special variant of a double-loop
pattern of laser stirring \cite{Allen_14} shows the effect similar to trap
shaking.

Ring dark solitons are known in optics \cite{Kivshar_15} and in Bose-Einstein
condensate \cite{Yukalov_8,Yukalov_9,Yukalov_10,Yukalov_11,Brand_16,Theocharis_17}.
Ring vortex solitons can be stable for weak atomic repulsion, but become
unstable for sufficiently strong effective repulsion, such as is typical
for trapped atoms. However, their lifetime can be rather long, of order of
seconds, which allows one to treat them as metastable objects
\cite{Car_18,Li_19}. Note that bright vortex solitons, existing for attractive
atomic interactions, are unstable in three dimensions
\cite{Adhikari_20,Adhikari_21}.

In the presence of external perturbations, vortex ring solitons experience
snake instability \cite{Kadomtsev_22} and can split into multiple fragments
\cite{Li_19}, and finally decay into vortex rings \cite{Anderson_23,Ginsberg_24}.
A reconnection between vortex lines can also lead to the emission of vortex
rings \cite{Svistunov_25,Laurie_26}.

In the present paper, we consider the situation, when Bose-condensed trapped
atoms are subject to an external perturbation, which corresponds to trap shaking.
The setup is explained in Sec. 2. Such trap shaking can generate different
nonlinear coherent modes, including vortices and ring vortex solitons
\cite{Yukalov_8,Yukalov_9,Yukalov_10,Yukalov_11}. The definition of coherent
modes, as stationary solutions to the NLS equation, is given in Sec. 3. When
the effective coupling parameter is asymptotically small, it is possible to
approximate the solutions to the NLS equation by Gauss-Laguerre modes. We show
that, even for arbitrarily strong coupling parameter, nonlinear coherent modes
can be well characterized by Hermite-Laguerre modes, provided the description
is done by means of optimized perturbation theory \cite{Yukalov_27,Yukalov_28}.

In our previous papers \cite{Shiozaki_29,Yukalov_30,Yukalov_31,Yukalov_32},
we have demonstrated that by shaking a trapped Bose-Einstein condensate, we
can generate the following nonequilibrium states: weak nonequilibrium,
vortex state, vortex turbulence, grain turbulence, and wave turbulence.
Analyzing now in more detail the initial stage of weak nonequilibrium, we
numerically simulate the three-dimensional NLS equation and find out that,
when no special resonance conditions are involved, in the regime of moderately
strong perturbations, vortex rings arise. This happens yet before vortices
start appearing in the vortex stage and far before vortex turbulence develops.
The results of the numerical simulation, observing vortex rings, are presented
in Sec. 4.

To summarize, in the present paper we show the following:

(i) We demonstrate that vortex ring solitons, that are a particular case 
of nonlinear coherent modes, can be characterized as Hermite-Laguerre modes, 
provided optimized perturbation theory is used. By vortex ring solitons, we 
understand circular dark solitons with circulation around the trap axis.  

(ii) A vortex ring is defined as a circular line of zero density, with 
circulation around each of its elements. By numerical simulations, we show 
that a moderate trap shaking produces vortex rings in a trapped $^{87}$Rb.
  
(iii) We explain why vortex rings arise before vortex ring solitons and even 
before vortices. This is because the generation of vortex rings requires much
less energy than that needed for the creation of other nonlinear coherent 
modes, including vortices and vortex ring solitons.

\section{Shaken trap setup}

We consider Bose atoms in a trap described by the harmonic trap potential
\be
\label{1}
U(\bfr) =  \frac{m}{2} \left ( \omega_\perp^2 r_\perp^2 + \om_z^2 r_z^2
\right ) \; ,
\ee
where $r_\perp^2 = r_x^2 + r_y^2$ is the radial spatial variable and $r_z$
is the longitudinal spatial variable. The trap aspect ratio is
\be
\label{2}
 \al \equiv \frac{\om_z}{\om_\perp} = \left ( \frac{l_\perp}{l_z}
\right )^2 \; .
\ee
Atoms interact through the local interaction potential
\be
\label{3}
 \Phi(\bfr ) = \Phi_0 \dlt(\bfr) \; , \qquad
\Phi_0 \equiv 4 \pi \hbar^2 \; \frac{a_s}{m} \; ,
\ee
where $m$ is atomic mass and $a_s$, scattering length. The strength of
interactions is characterized by the dimensionless coupling parameter
\be
\label{4}
  g \equiv 4 \pi N \; \frac{a_s}{l_\perp} \; ,
\ee
where $N$ is the number of atoms.

Temperature is assumed to be low, so that almost all atoms are in the Bose
condensed state. The shape of the atomic cloud in the trap can be estimated
in the Thomas-Fermi approximation \cite{Pethick_7}, which yields the cloud
radius
\be
\label{5}
 R = 1.036 \; l_\perp (\al g)^{1/5}
\ee
and the cloud length
\be
\label{6}
L = \frac{2R}{\al} = 2.072 \; \frac{l_\perp}{\al} \; ( \al g)^{1/5} \; .
\ee
Hence the cloud volume is
\be
\label{7}
V = \pi R^2 L = 6.986 \; \frac{l_\perp^3}{\al} \; (\al g)^{3/5} \; .
\ee

Trap shaking is accomplished by imposing an additional time-dependent
potential, resulting in the total potential
\be
\label{8}
 U(\bfr,t) = U(\bfr) + V(\bfr,t)
\ee
consisting of the stationary confining potential $U(\bfr)$ and an alternating
potential $V(\bfr,t)$. The latter is chosen so that to vibrate the trap
without rotation. The shaking potential has the form
\be
\label{9}
 V(\bfr,t) =
\frac{m}{2} \; \Om_x^2(t) ( x' - x_0')^2 +
\frac{m}{2} \; \Om_y^2(t) (y' - y'_0 )^2 +
\frac{m}{2} \; \Om_z^2(t) (z' - z'_0 )^2 \; ,
\ee
with the vibration frequencies
\be
\label{10}
 \Om_\al(t) = A_\al \om_\al [ 1 - \cos(\om t) ] \; .
\ee
The primed variables $x'$, $y'$, and $z'$ are shifted and tilted with
respect to the original spatial variables $x$, $y$, and $z$, as is explained
in Ref. \cite{Seman_33}.

\section{Vortex ring solitons}

Vortex ring solitons are a particular case of nonlinear coherent modes that
can be introduced as solutions to the NLS equation possessing a circulation
around the trap axis, characterized by a nonzero winding number. Coherent 
modes can be generated by strongly shaking a trap. In a cylindrical trap, 
coherent modes can be represented as Hermite-Laguerre modes, which can be 
done not only for weak interactions, but for any strong atomic interactions, 
provided {\it Optimized Perturbation Theory} \cite{Yukalov_27,Yukalov_28} 
is employed.

Stationary states of Bose-condensed atoms are described by the stationary NLS
equation
\be
\label{11}
 \hat H[\eta ] \eta(\bfr) = E \eta(\bfr) \; ,
\ee
with the nonlinear Hamiltonian
\be
\label{12}
 \hat H[\eta] = - \; \frac{\hbar^2\nabla^2}{2m} + U(\bfr) +
\Phi_0 | \; \eta(\bfr) \; |^2 \; .
\ee
The normalization condition for the condensate function is
\be
\label{13}
\int | \; \eta(\bfr) \; |^2 d\bfr = N \; .
\ee

It is convenient to introduce the dimensionless spatial variables
\be
\label{14}
r \equiv \frac{r_\perp}{l_\perp} \; , \qquad
z \equiv \frac{r_z}{l_\perp}
\ee
and the dimensionless condensate function
\be
\label{15}
\eta(\bfr) \equiv \sqrt{ \frac{N}{l_\perp^3} } \; \psi(r,\vp,z) \; .
\ee
Then, with the dimensionless Hamiltonian
\be
\label{16}
\hat H [\psi ] \equiv \frac{\hat H[\eta]}{\hbar\om_\perp} \; ,
\ee
the eigenvalue problem (\ref{11}) reads as
\be
\label{17}
\hat H[\psi_{nmj} ] \psi_{nmj}(r,\vp,z) = E_{nmj} \psi_{nmj}(r,\vp,z) \; .
\ee
Here
\be
\label{18}
\hat H[\psi] \equiv - \; \frac{1}{2} \; \nabla^2 + \frac{1}{2} \left (
r^2 + \al^2 z^2 \right ) + g | \; \psi \; |^2 \; ,
\ee
with
$$
\nabla^2 = \frac{\prt^2}{\prt r^2} + \frac{1}{r}\; \frac{\prt}{\prt r}
+ \frac{1}{r^2}\; \frac{\prt^2}{\prt \vp^2} + \frac{\prt^2}{\prt z^2} \; .
$$
The normalization condition (\ref{13}) becomes
\be
\label{19}
\int_0^\infty r \; dr \; \int_0^{2\pi} d\vp \;
\int_{-\infty}^{\infty} | \; \psi_{nmj}(r,\vp,z) \; |^2 \; dz = 1 \; .
\ee

In optimized perturbation theory \cite{Yukalov_27,Yukalov_28}, we start
with a trial Hamiltonian
\be
\label{20}
\hat H_0 = - \; \frac{1}{2} \; \nabla^2 + \frac{1}{2} \left ( u^2 r^2
+ v^2 z^2 \right )
\ee
including control functions $u$ and $v$ to be defined from an optimization
condition. The trial wave function, corresponding to Hamiltonian (\ref{20}),
is given by the expression
$$
\psi_{nmj}(r,\vp,z)  = \left [ \; \frac{2n! u^{|m|+1}}{(n+|m|)!} \;
\right ]^{1/2}  r^{|m|} \;
\exp\left ( -\; \frac{u}{2} \; r^2 \right )\; L_n^{|m|}(ur^2) \;
\frac{e^{im\vp}}{\sqrt{2\pi}} \; \times
$$
\be
\label{21}
\times \left ( \frac{v}{\pi} \right )^{1/4} \frac{1}{\sqrt{2^j \; j!}} \;
\exp \left ( - \; \frac{v}{2} \; z^2 \right ) H_j ( \sqrt{v} \; z) \; ,
\ee
where $L_n^m$ is a generalized Laguerre polynomial and $H_j$ is a Hermite
polynomial. Here $ n = 0,1,2,\ldots$ is a radial quantum number,
$m=0,\pm 1, \pm 2,\ldots$ is an azimuthal quantum number or the winding
number of circulation, and $j=0,1,2,\ldots$ is an axial quantum number.

The optimization condition, defining the control functions, can be chosen
in several forms, the simplest of which is
\be
\label{22}
\left ( \dlt u \; \frac{\prt}{\prt u} + \dlt v \; \frac{\prt}{\prt v}
\right ) \left ( \psi_{nmj} , \; \hat H[\psi_{nmj} ] \psi_{nmj} \right )
= \left ( \dlt u \; \frac{\prt}{\prt u} + \dlt v \; \frac{\prt}{\prt v}
\right ) E_{nmj} = 0 \; .
\ee
The spectrum of coherent modes is
\be
\label{23}
E_{nmj} = \frac{p}{2} \left ( u + \frac{1}{u} \right ) +
\frac{q}{4} \left ( v + \frac{\al^2}{v} \right ) +
\frac{u\sqrt{v}}{(2\pi)^{3/2}} \; J_{nmj} g \; ,
\ee
in which
\be
\label{24}
p \equiv p_{nm} = 2n + | m| + 1 \; , \qquad q \equiv q_j = 2j +1
\ee
and
\be
\label{25}
J_{nmj} \equiv \frac{(2\pi)^{3/2}}{u\sqrt{v} }
\int | \; \psi_{nmj}(r,\vp,z) \; |^4 r dr d\vp dz \; .
\ee

The optimization condition (\ref{22}) gives the equations for the
control-functions:
\be
\label{26}
p \left ( 1 -\; \frac{1}{u^2} \right ) + \frac{G}{\al p} \; \sqrt{ \frac{v}{q} } = 0 \; ,
\qquad
q \left ( 1 -\; \frac{\al^2}{v^2} \right ) + \frac{uG}{\al p\sqrt{vq}} = 0 \; ,
\ee
where we define the effective coupling parameter
\be
\label{27}
G \equiv G_{nmj} = \frac{2p\sqrt{q}}{(2\pi)^{3/2} } \; J_{nmj} \al g \; .
\ee

In a typical situation with trapped atoms, the effective coupling is strong,
such that
\be
\label{28}
g \gg 1 \; , \qquad G \gg 1 \; .
\ee
Then the control functions are
\be
\label{29}
u \simeq \frac{p}{G^{2/5}} \; , \qquad v \simeq \frac{\al^2 q}{G^{2/5}} \; .
\ee

We are interested in the radial shape of the condensate density. To this end,
we set $j=0$ and define the radial density
$$
\rho_{nm}(r) \equiv \int | \; \psi_{nm0}(r,\vp,z)\; |^2 d\vp dz \; ,
$$
for which we obtain
\be
\label{30}
\rho_{nm}(r) = \frac{2n! u^{m+1}}{(n+m)!} \; r^{2m} \exp \left ( - u r^2 \right )
\left [ L_n^m\left (ur^2 \right ) \right ]^2 \; ,
\ee
where $m\geq 0$.

The effective coupling parameter is
$$
G = \frac{2p}{(2\pi)^{3/2}} \; J_{nm} \al g \qquad ( q = 1) \; ,
$$
with
$$
J_{nm} \equiv \frac{1}{u} \int_0^\infty \rho_{nm}^2(r) \; r dr \; .
$$
In particular,
$$
J_{00} = 1 \; , \qquad J_{01} = \frac{1}{2} \; , \qquad J_{02} = \frac{3}{8} \; ,
$$
$$
J_{10} = \frac{1}{2} \; , \qquad J_{11} = \frac{5}{16} \; , \qquad
J_{12} = \frac{1}{4} \; ,
$$
$$
J_{20} = \frac{11}{32} \; , \qquad J_{21} = \frac{15}{64} \; , \qquad
J_{22} = \frac{199}{1024} \; .
$$

In Figs. $1$, $2$, and $3$, we present the radial density of coherent modes
for different quantum numbers. The radial number $n$ defines the number
of nodes, while the winding number $m$ characterizes circulation. For numerical
calculations, we accept the parameters as in our previous experiments with
$^{87}$Rb, described in Refs.
\cite{Shiozaki_29,Yukalov_30,Yukalov_31,Yukalov_32,Seman_33}, where
$$
\al = 0.11 \; , \qquad g = 1.96\times 10^4 \; , \qquad
\al g = 2.156 \times 10^3 \; .
$$
For comparison, we present vortex solutions, with nonzero $m$, and
non-circulating solutions, with $m = 0$.

\section{Generation of vortex rings}

The typical radii of rings appearing in a trap are much larger than the
coherence length $l_c$ and, clearly, smaller than the trap radius $R$,
\be
\label{31}
r_R \gg l_c \; , \qquad r_R < R \; .
\ee
The coherence length
\be
\label{32}
l_c \equiv \frac{\hbar}{mc} = \frac{1}{\sqrt{4\pi\rho a_s}} \qquad
\left ( c = \frac{\hbar}{m} \; \sqrt{4\pi\rho a_s} \right )
\ee
defines the vortex ring core.

Ring characteristics can be estimated resorting to the Thomas-Fermi
approximation \cite{Pethick_7}. Normalizing the Thomas-Fermi condensate
function yields the condensate energy
$$
E_{TF} = \frac{1}{2} \left ( \frac{15}{4\pi} \; \al g \right )^{2/5} =
0.5367 (\al g)^{2/5}
$$
expressed in units of $\hbar \om_\perp$. The condensate density at the trap
center reads as
$$
\rho(0) = 0.537 \; \frac{\al N}{l_\perp^3 (\al g)^{3/5} } \; .
$$
Taking the latter for estimating the coherence length gives
\be
\label{33}
l_c = R \; \sqrt{ \frac{4\pi L}{15 g l_\perp} } = 1.365 \; \frac{l_\perp}{(\al g)^{1/5} } \; .
\ee

The vortex ring characteristics have been considered by Iordanskii
\cite{Iordanskii_34}, Amit and Gross \cite{Amit_35}, Roberts and Grant
\cite{Roberts_36}, and Jones and Roberts \cite{Jones_37}. Following
Refs. \cite{Amit_35,Roberts_36}, for the energy of a vortex ring, in units
of $\hbar \om_\perp$, we have
\be
\label{34}
E_R = 2\pi^2 \rho r_R l_\perp^2 \; \ln \left ( 2.25 \; \frac{r_R}{l_c} \right ) \; ,
\ee
where the density $\rho$ is to be taken at the location of the ring.

Since the density is smaller close the boundary of a trapped atomic cloud,
we expect that it is easier to create rings close to the boundary, since
this requires smaller energy. The following numerical simulations confirm this.

Generally, a vortex ring moves with velocity that, far from the cloud edges,
can be estimated as
\be
\label{35}
v_R =  \frac{\hbar}{2m r_R} \;
\ln \left ( 6.12 \; \frac{r_R}{l_c} \right ) \; ,
\ee
the velocity being directed perpendicular to the ring plane. The ring
possesses a stationary position (although nonequilibrium) when its radius
is $r_R \sim 0.6 \; R$. But in the standard situation the ring has an
inherent tendency to propagate along the $z$ axis. In a trap, the ring
oscillates \cite{Jackson_38,Reichl_39}. Estimating the period of the ring
oscillation \cite{Bulgac_40} one has
\be
\label{36}
T_R \sim \frac{4 \pi R/l_c}{\omega_z \sqrt{\ln(R/l_c)}} \; .
\ee
Substituting here the expressions for $R$ and $l_c$, we find
\be
\label{37}
T_R \sim \frac{4\pi(\alpha g)^{2/5}}{\omega_z \sqrt{\ln(0.5\alpha g)}} \; .
\ee
For the accepted setup, where $\omega_z = 2 \pi \times 23$ Hz, we get
$T_R \sim 0.7$s.

We accomplish numerical solution of the NLS equation, with the shaking
potential (\ref{9}), using the method of Ref. \cite{Novikov_41}. It turns
out that the first nonequilibrium stage of perturbation, classified in
Refs. \cite{Yukalov_30,Yukalov_31,Yukalov_32} as the regime of weak
nonequilibrium, is not as dull. At this stage, there are yet no vortices.
However, there appear vortex rings. A vortex ring is defined as a circular
line of zero density, with a winding number $\pm 1$ around each of the
line elements. Numerically, zero density implies $10^{-8}\rho(0)$. Some
typical vortex rings are shown in Fig. 4. Rings usually appear in pairs,
so that their total number is even.

In order to understand why vortex rings appear before vortices, we need
to compare their energies. The energy of a vortex of length $L$ can be
written \cite{Pethick_7,Yukalov_31,Yukalov_32} as
\be
\label{38}
 E_L = \frac{2\pi}{3} \; \rho(0) l_\perp^2 L
\ln \left ( 0.949 \; \frac{R}{l_c} \right )  \; ,
\ee
where $\rho(0)$ is the density at the trap center. Since the total number
of atoms, in the Thomas-Fermi approximation, is
\be
\label{39}
 N \equiv \int \rho(\bfr) \; d\bfr = \frac{4\pi}{15}\; \rho(0) R^2 L \;  ,
\ee
it is also possible to write
$$
 E_L = \frac{5l^2_\perp}{2R^2} \; N
\ln \left ( 0.949 \; \frac{R}{l_c} \right )  \;   .
$$

Comparing the vortex energy (\ref{38}) with the ring energy (\ref{34}),
we take $r_R \sim 0.6 R$, which gives the relation
$\rho(r_R) \sim 0.7 \rho(0)$ between the atomic density close to the ring
location and the density at the center. Then we obtain
\be
\label{40}
 \frac{E_R}{E_L} \sim \alpha \;  .
\ee
In the considered setup, $\alpha \sim 0.1$. Hence the ring energy is an
order smaller than the vortex energy. Therefore it is much easier to
generate vortex rings, requiring to pump into the system much less energy
than that needed for producing vortices. It is possible to expect that
up to around ten vortex rings could be generated before vortices will
start arising. The energy of other coherent modes, including vortex ring 
solitons is larger than that of basis vortices with the winding number 
$\pm 1$. Hence their appearance is even less probable than that of
the simple vortices, unless special resonant conditions are imposed.

\section{Conclusion}

We have analyzed the possibility of creating vortex ring solitons and
vortex rings in a shaken Bose-Einstein condensate of trapped atoms.
The consideration is based on the NLS equation, assuming that practically
all atoms are condensed. Vortex ring solitons are dark solitonic
solutions, with circulation around the symmetry axis of the trap. Vortex
rings are circular vortices with circulation around each of their elements.
Vortex ring solitons can be described analytically by solving the NLS
equation using optimized perturbation theory. This solution is valid for
an arbitrary strength of atomic interactions, contrary to the simple
perturbation theory admissible for only very weak interactions.

Numerical simulation of the three-dimensional NLS equation demonstrates
the appearance of vortex rings. This happens in the regime of a rather
moderate perturbation, yet before vortices could arise. That regime
corresponds to the weak nonequilibrium stage, according to the
classification of Refs. \cite{Yukalov_30,Yukalov_31,Yukalov_32}. 

The energy of a vortex ring is essentially smaller than that of a basic
vortex with circulation $\pm 1$. In turn, the energy of a simple vortex is 
smaller than that of other coherent modes, including vortex ring solitons
\cite{Courteille_42,Bagnato_43}. This explains why, under moderate trap 
shaking, vortex rings appear long before vortices, since their generation 
requires much less energy to be injected into the trap. Even if 
higher-energy coherent modes would be produced at the initial moment of 
time, they would disintegrate into the lower-energy modes, such as 
vortices and vortex rings. Vortex rings can also occur in the stage of 
vortex turbulence \cite{Tsubota_44,Nemirovskii_45}. However, to reach this 
regime requires much stronger trap shaking. The investigation of the 
possible vortex-ring creation under vortex turbulence, realized by trap 
shaking, is in the progress.

\section*{Acknowledgement}

Financial support from the Russian Foundation for Basic Research
(grant $\#$ 14-02-00723) and from the University of S\~ao Paulo 
NAP Program is appreciated.

\begin{figure}[H]
\centering
\includegraphics[width=6cm]{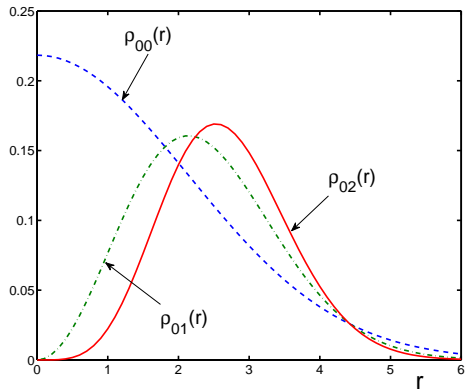}
\caption{Radial density $\rho_{0m}(r)$ as a function of the dimensionless radius
(in units of the transverse oscillator length $l_\perp$). The density of the
ground state $\rho_{00}(r)$ (dashed line) is compared with the densities of
the vortex states with the winding number $m=1$ (dashed-dotted line) and
$m=2$ (solid line).}
\label{Fig1}
\end{figure}

\begin{figure}[H]
\centering
\includegraphics[width=6cm]{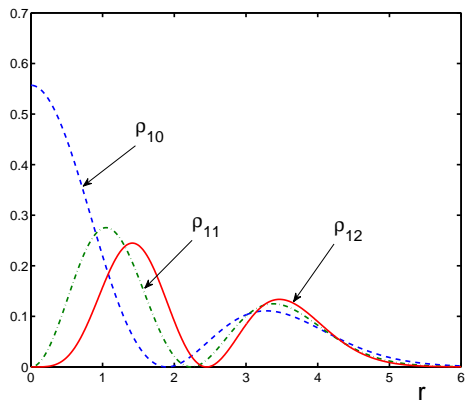}
\caption{Radial density $\rho_{1m}(r)$ of the ring state
$\rho_{10}(r)$ (dashed line), composite vortex $+$ ring
states $\rho_{11}(r)$ (dashed-dotted line) and $\rho_{12}(r)$ (solid line).
}
\label{Fig2}
\end{figure}

\begin{figure}[H]
\centering
\includegraphics[width=6cm]{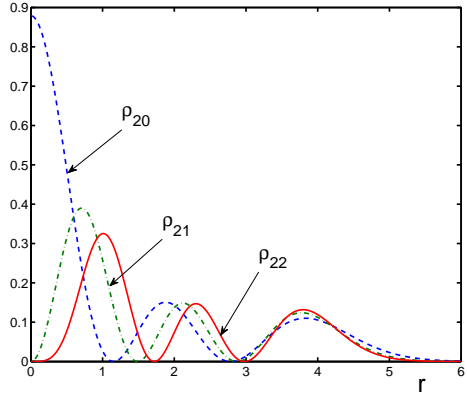}
\caption{Radial density $\rho_{2m}(r)$ of the double ring state
$\rho_{20}(r)$ (dashed line), composite vortex $+$ double-ring
states $\rho_{21}(r)$ (dashed-dotted line) and $\rho_{22}(r)$ (solid line).}
\label{Fig3}
\end{figure}

\begin{figure}[H]
\centering
\includegraphics[width=14.5cm]{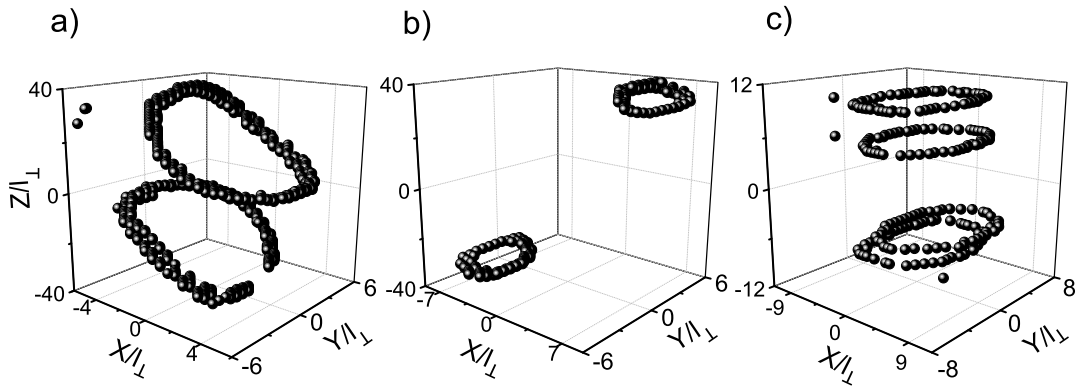}
\caption{Spatial location of vortex rings after different shaking time:
(a) $t=10.5$ ms; \newline (b) $t=11$ ms; (c) $t=14.4$ ms.}
\label{Fig4}
\end{figure}

\end{document}